\documentclass[aps,prl,reprint,superscriptaddress,floatfix]{revtex4-2}

\usepackage{graphicx}
\usepackage{amsmath,amssymb,bm}
\usepackage{microtype}
\usepackage{booktabs}
\usepackage{siunitx}
\usepackage{xcolor}
\usepackage{hyperref}
\hypersetup{
  hidelinks,
  pdftitle={Symmetry-Enabled Ferroelectric Valley Transduction in
            Rhombohedral Bilayer WSe2},
  pdfauthor={Katsunori Wakabayashi, Souren Adhikary, Tomoaki Kameda,
             Kazuhito Tsukagoshi},
  pdfsubject={Valley physics of R-stacked bilayer WSe2},
  pdfkeywords={WSe2, R stacking, sliding ferroelectricity, K-Gamma valley,
               transconductance}}

\newcommand{\tK}{t_K^{\mathrm R}}
\newcommand{\tG}{t_\Gamma^{\mathrm R}}
\newcommand{\Dsym}{D^{*}}

\newcommand{\VPexpt}{56}                
\newcommand{\VPexptsd}{3}               
\newcommand{\Dbuiltin}{0.102}           
\newcommand{\UFEdft}{83.43}                
\newcommand{\polshiftdft}{41.7}           
\newcommand{\UFEhalfdft}{41.7}         
\newcommand{\tKlimitcanon}{0.017}        
\newcommand{\tGRband}{361.8}             
\newcommand{\dratioGK}{0.76}            
\newcommand{\UratioGK}{0.79}            
\newcommand{\gdnullpct}{1.8}            
\newcommand{\gdnulllow}{0.2}            
\newcommand{\mulayerslope}{1.84}         

\newcommand{\tslopeKM}{160}              
\newcommand{\tslopeKG}{123}              
\newcommand{\tinterKM}{-0.06}            
\newcommand{\tinterKG}{+0.13}            

\begin{document}

\title{Symmetry-Enabled Ferroelectric Valley Transduction in Rhombohedral
Bilayer \texorpdfstring{WSe$_2$}{WSe2}}

\author{Katsunori Wakabayashi}
\email{WAKABAYASHI.Katsunori@nims.go.jp}
\affiliation{Research Center for Materials Nanoarchitectonics (MANA), National Institute for Materials Science (NIMS), Tsukuba 305-0044, Japan}
\author{Souren Adhikary}
\affiliation{Research Center for Materials Nanoarchitectonics (MANA), National Institute for Materials Science (NIMS), Tsukuba 305-0044, Japan}
\author{Tomoaki Kameda}
\affiliation{Research Center for Materials Nanoarchitectonics (MANA), National Institute for Materials Science (NIMS), Tsukuba 305-0044, Japan}
\author{Kazuhito Tsukagoshi}
\affiliation{Research Center for Materials Nanoarchitectonics (MANA), National Institute for Materials Science (NIMS), Tsukuba 305-0044, Japan}

\date{\today}

\begin{abstract}
Ferroelectric polarization couples to the layer degree of freedom in a bilayer, but a valley-selective energy response requires the valleys to carry different interlayer character. Rhombohedral bilayer WSe$_2$ meets that condition by symmetry: three-fold rotation forbids interlayer tunneling at the $K$ valley, while strong hybridization remains at $\Gamma$. The resulting contrast converts the band-effective ferroelectric layer asymmetry into a shift of the $K$ edge, and so displaces the $K$--$\Gamma$ landscape: a finite-field fit that does not impose the rule leaves the $K$ coupling consistent with zero, and the shift reaches \polshiftdft~meV, the most that asymmetry can give a single valley. Reversing the polarization mirrors the valley landscape in displacement field, so the two registries have equal and opposite differential transconductance $g_d$ at zero field. In the ideal symmetric limit a second parity holds about each registry's own compensation field and places a zero of $g_d$ there. The valley-dependent screening calculated here displaces that zero by only \gdnulllow\% at the read point, so device asymmetry limits the accuracy. The result links a known stacking selection rule to ferroelectric valley transduction, with a quantitative valley shift and a directly testable transport signature.
\end{abstract}

\maketitle

A ferroelectric polarization is nonvolatile, and a valley index labels electronic states in momentum
space~\cite{Schaibley2016_valleytronics,Xiao2012_valley}.  Layered transition-metal dichalcogenides host
both, and stacking controls them~\cite{Manzeli2017_TMDreview}.  Reversing the polarization reverses a
layer-dependent potential, but two valleys can share the same layer character, and such a potential then
moves them together.  The mechanism has to make one layer potential act differently on different points of
the Brillouin zone.  We call the resulting conversion of a polar layer asymmetry into a relative shift
between valleys of different interlayer character \emph{ferroelectric valley transduction}.

\begin{figure*}[t]
  \centering
  \includegraphics[width=0.94\textwidth]{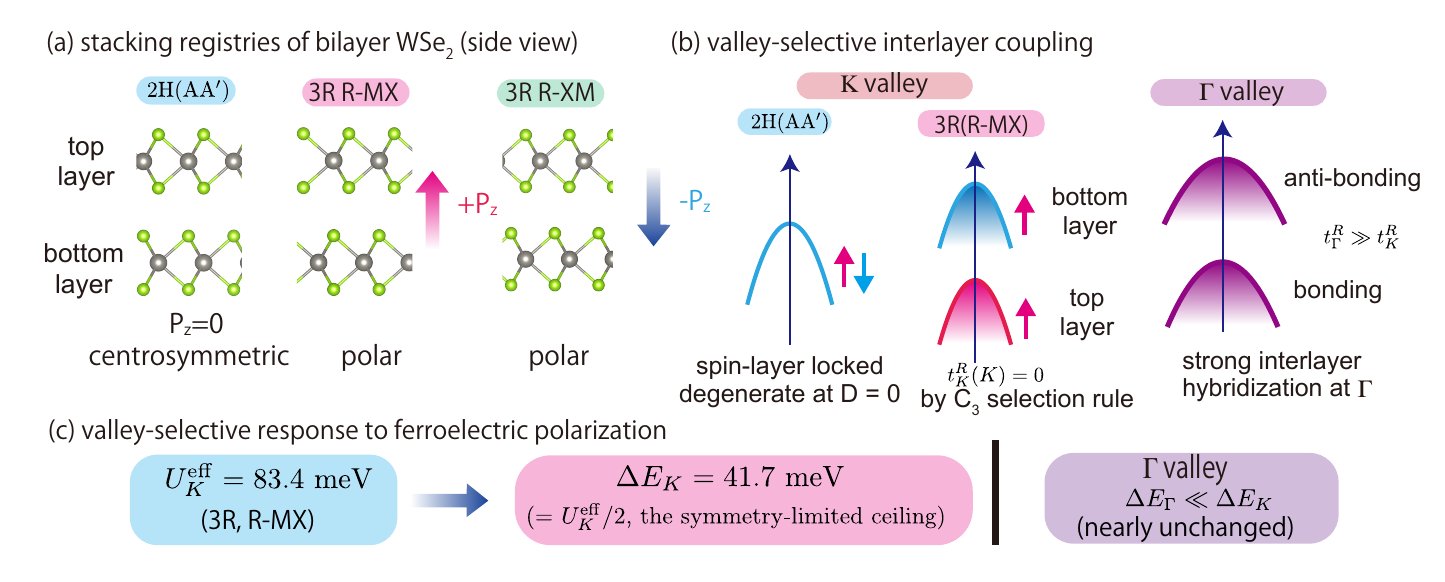}
  \caption{\textbf{Stacking symmetry, valley-selective coupling, and what the
  polarization does with them.}
  (a) The centrosymmetric 2H bilayer and the two rhombohedral registries,
  which an interlayer slide takes into one another and which have opposite
  out-of-plane polarizations.
  (b) The two valleys meet the layer degree of freedom differently. At $K$ in
  2H the branches are degenerate and opposite spins forbid the
  spin-conserving hopping, while the orbital matrix element stays
  finite~\cite{Jones2014_spinlayer,Riley2014_WSe2spinvalley}; the states
  therefore stay on their layers. At $K$ in R stacking the two layer states
  have the \emph{same} spin, and the three-fold selection rule forbids the
  hopping instead, $\tK(K)=0$. At $\Gamma$ the layers hybridize into bonding and
  antibonding combinations in either stacking. Color denotes the layer
  within each $K$ panel, bottom in blue and top in red.
  (c) The consequence. The polarization's interlayer step is converted into a
  shift of the $K$ edge, the most a decoupled valley can take, while
  $\Gamma$, held together by $t_\Gamma$, barely moves. Panel~(b) is drawn for
  R-MX, whose $K$ edge lies on the bottom layer; R-XM is its layer-exchanged
  mirror. Figure~\ref{fig:landscape} makes the field dependence
  quantitative.}
  \label{fig:concept}
\end{figure*}

Rhombohedral transition-metal dichalcogenide bilayers provide such a mechanism. Their two parallel-oriented layers can be translated into two polar registries with opposite out-of-plane polarization [Fig.~\ref{fig:concept}(a)], predicted in sliding bilayers~\cite{LiWu2017_sliding} and since observed across layered and marginally twisted systems~\cite{Fei2018_WTe2,Weston2022_marginal} and in parallel-stacked bilayer WSe$_2$ itself, whose built-in interlayer potential is measured at \VPexpt(\VPexptsd)~meV~\cite{Wang2022_interfacialFE}, where the two registries meet at mobile domain walls~\cite{Yang2024_domainwall}. For the $K$ valley, the interlayer matrix element of a 3R bilayer vanishes by a three-fold-rotation selection rule \cite{Akashi2015_3RMoS2}; at $\Gamma$, by contrast, the valence states have strong out-of-plane orbital character and form bonding--antibonding combinations [Fig.~\ref{fig:concept}(b)]. The selection rule itself is known.  This Letter applies the rule to a competing-valley landscape in a bilayer that is also ferroelectric, and reports the electrical signature of the transduction that follows.

Bilayer WSe$_2$ is a particularly direct realization because the valence maxima at $K$ and $\Gamma$ lie close enough that their relative position controls the hole population and transport~\cite{Movva2018_valleyPRL,Li2026_RWSe2}. The maximum moves from $K$ towards $\Gamma$ as layers are added, because the layers hybridize~\cite{Komsa2012_indirect,Zhang2014_MoSe2,Yeh2015_WSe2muARPES}. Which of the two wins in the bilayer is unsettled: an optical study assigns the edge to $K$~\cite{Li2026_RWSe2} and photoemission on quasi-free-standing rhombohedral bilayers to $\Gamma$~\cite{Chapuis2026_3RW110}. A third photoemission study resolves both pockets, and the accompanying calculations reverse the ordering between a hybrid functional and a quasiparticle correction~\cite{Mahmoudi2023_3Rbilayer}. Every claim below is a \emph{shift} of the landscape or a parity of the response, and holds under either assignment.  The two valleys differ sharply in mass~\cite{Kormanyos2015_kp,Fallahazad2016_mass,Chapuis2026_3RW110}, and moving the landscape changes the current. Two companion studies treat that landscape as fixed and move the chemical potential across it~\cite{PaperI_valleyGAA,PaperII_transconductance}; here the landscape itself moves. We show that the contrast between a symmetry-decoupled $K$ valley and a strongly hybridized $\Gamma$ valley converts the built-in polar layer asymmetry almost entirely into a shift of the $K$--$\Gamma$ landscape. The two polar registries then become mirror images along the displacement-field axis, and this mirror relation
generates a differential-transport parity.  That parity reads the stored polarization through the sign, or through the field position of a zero, and not through an absolute current difference.

For each valley we use a minimal two-layer Hamiltonian in the top/bottom layer basis.
\begin{equation}
H_\nu^{(s)}(D)=\bar E_\nu I+t_\nu\tau_x+\frac{\Delta_\nu^{(s)}(D)}{2}\tau_z,
\label{eq:H}
\end{equation}
with
\begin{equation}
\Delta_\nu^{(s)}(D)=sU^\nu+e d_\nu D+U_{\rm env}.
\label{eq:Delta}
\end{equation}
Here $s=-1$ for R-MX and $s=+1$ for R-XM, $d_\nu$ is the screened dipole length, and $U_{\rm env}$ represents a static top--bottom environmental asymmetry. The mirror relation below requires $U_{\rm env}=0$. A built-in asymmetry shifts both compensation fields one way, whereas the polarization shifts them oppositely.  The pair then ceases to sit symmetrically about zero field and the relation is broken. Equation~(\ref{eq:Delta}) lets both $U^\nu$ and $d_\nu$ depend on the valley. The ideal limit treated below takes the two valleys to share a compensation field, $U^K/d_K=U^\Gamma/d_\Gamma$, which is weaker than requiring either quantity to be common. The direct WSe$_2$ calculations measure the departure from that condition, and we quantify its effect when we reach the readout zero.

The upper branch of Eq.~(\ref{eq:H}) is
\begin{equation*}
E_{\nu,+}=\bar E_\nu+\sqrt{t_\nu^2+(\Delta_\nu/2)^2},
\end{equation*}
and its susceptibility to a layer asymmetry is
\begin{equation}
\frac{\partial E_{\nu,+}}{\partial\Delta_\nu}
=\frac12\frac{\Delta_\nu/2}{\sqrt{t_\nu^2+(\Delta_\nu/2)^2}}.
\label{eq:susc}
\end{equation}
Thus a layer-decoupled valley reaches the largest slope Eq.~(\ref{eq:susc}) allows, $|\partial E/\partial\Delta|=1/2$, whereas a strongly hybridized valley responds only weakly. The R-stacked $K$ valley is an exact realization of the first limit at the high-symmetry point. If $|t\rangle$ and $|b\rangle$ are the layer Bloch states at $K$, the lateral registry gives them different $C_3$ eigenvalues $e^{i\phi_t}$ and $e^{i\phi_b}$. Since $[H,C_3]=0$,
\begin{equation*}
\langle t|H|b\rangle
=e^{i(\phi_b-\phi_t)}\langle t|H|b\rangle,
\end{equation*}
and $\phi_b-\phi_t=\pm2\pi/3$ forces the matrix element to vanish \cite{Akashi2015_3RMoS2}. The vanishing is an interference condition on the Bloch phases, not a small material parameter.

\begin{figure*}[t]
  \centering
  \includegraphics[width=0.94\textwidth]{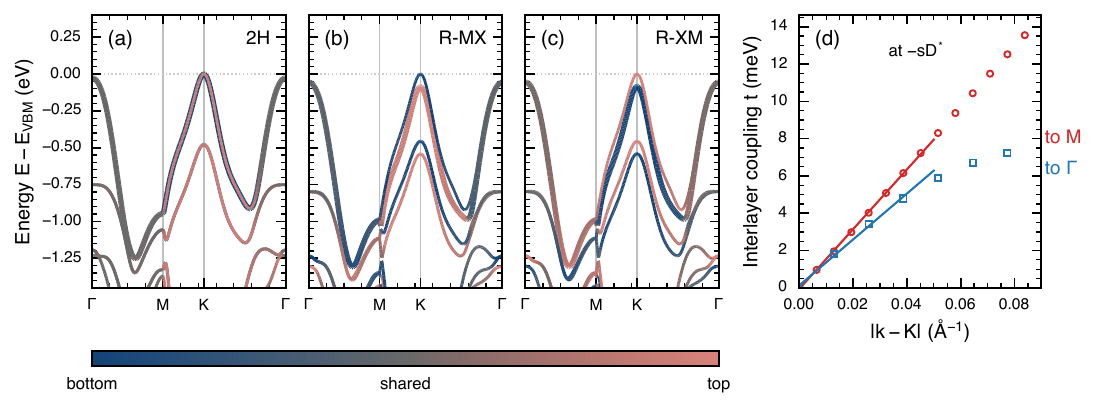}
  \caption{\textbf{Microscopic origin of the valley-selective response.}
  (a)--(c) Layer-resolved valence bands of 2H, R-MX, and R-XM at zero field. Color denotes the layer weight: the R-stacked $K$ edge is layer polarized, whereas the valence states at $\Gamma$ are strongly shared between the layers.
  (d) At the compensation field the interlayer coupling extracted from the two-level inversion vanishes linearly at $K$ and recovers away from it. Lines are fits over $|\mathbf{k}-K|\le0.05$~\AA$^{-1}$; separate fits toward $M$ and $\Gamma$ give slopes \tslopeKM{} and \tslopeKG~meV\,\AA{}, with intercepts $\tinterKM$ and $\tinterKG$~meV, respectively.}
  \label{fig:micro}
\end{figure*}

\emph{Calculations.}---VASP~\cite{Kresse1996_vasp} with projector augmented-wave potentials~\cite{Blochl1994_paw,KresseJoubert1999_paw}, the PBE functional~\cite{Perdew1996_pbe}, non-collinear spin--orbit coupling and DFT-D3 dispersion with Becke--Johnson damping~\cite{Grimme2010_d3,Grimme2011_bj}; 600~eV cutoff, $10^{-6}$~eV threshold, 30~\AA{} cell, non-spherical gradients inside the augmentation spheres and one-center occupancies mixed to $l=4$. Positions were relaxed at fixed cell to $5\times10^{-4}$~eV\,\AA$^{-1}$ on a $\Gamma$-centered $24\times24\times1$ mesh~\cite{MonkhorstPack1976}. The states reported here are evaluated on that mesh in the polar registries and on $12\times12\times1$ in the nonpolar one, where the charge-density symmetrization is disabled; both meshes contain $K$ exactly. A finite field enters as a sawtooth potential along the stacking axis; the polar R-stacked slabs use the dipole correction~\cite{NeugebauerScheffler1992_dipole}. Convergence tests, the parameter extraction and its conditioning, and a hybrid-functional check are in the companion Article~\cite{CompanionArticle}.

Our first-principles calculations make that distinction visible. At zero field the $K$ states of both R registries are almost completely layer polarized, whereas the top valence states at $\Gamma$ are shared between the layers [Fig.~\ref{fig:micro}(a)--(c)]. The zone-center bonding--antibonding splitting gives $\tG\simeq\tGRband$~meV. An unconstrained finite-field scan, which assumes nothing about the symmetry, puts $\tK$ below $\tKlimitcanon$~meV --- three orders of magnitude under $k_BT$ at room temperature, and at the resolution of the calculation and not at any scale of the crystal. More decisively, at the compensation field, where the two $K$-layer on-site energies coincide, the inferred coupling rises linearly away from $K$ with slopes \tslopeKM{} and \tslopeKG~meV\,\AA{} along the two inequivalent directions, while both extrapolated intercepts are numerically zero [Fig.~\ref{fig:micro}(d)]. A merely small coupling would remain small in a neighborhood of $K$; the observed linear node is the signature of the selection rule.

The ferroelectric consequence follows directly from Eq.~(\ref{eq:susc}). The band-effective asymmetry of the R-stacked $K$ doublet is $U_K^{\rm eff}=\UFEdft$~meV. In the symmetry limit $t_K(K)=0$, switching on this asymmetry shifts the upper $K$ edge by the maximum allowed amount $U_K^{\rm eff}/2=\UFEhalfdft$~meV. The unconstrained finite-field fit, which does not assume the selection rule, returns a coupling consistent with zero and gives the same \polshiftdft~meV [Fig.~\ref{fig:concept}(c)]. The two routes agree without either granting the other's premise. The $\Gamma$ edge barely responds because $t_\Gamma\gg U^\Gamma$.

\begin{figure*}[t]
  \centering
  \includegraphics[width=0.94\textwidth]{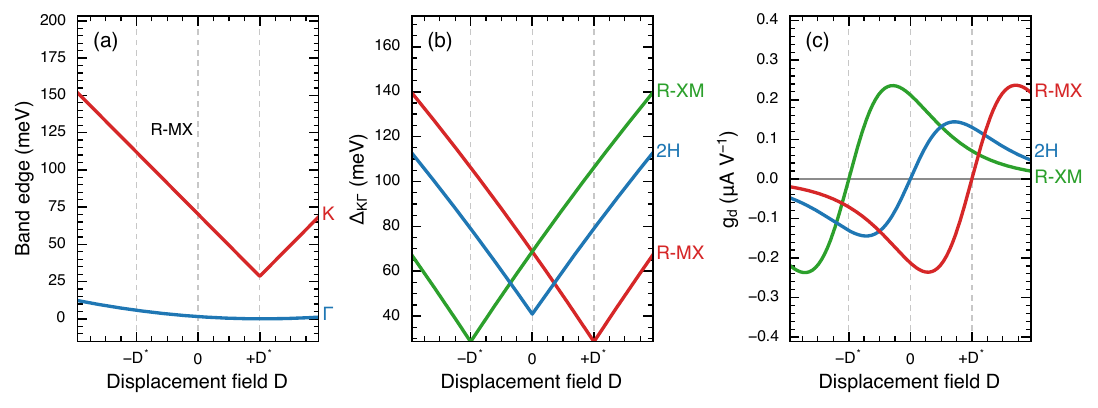}
  \caption{\textbf{From transduction to a parity readout.} The same three
  fields $-D^{*}$, $0$ and $+D^{*}$ are marked in every panel.
  (a) For R-MX, the $K$ edge follows the layer asymmetry at close to the
  ceiling of Eq.~(\ref{eq:susc}) while $\Gamma$ barely moves; color
  denotes the valley in this panel and the stacking in the other two.
  (b) The two polar landscapes are exact field-axis mirrors of one another,
  with minima at $+D^{*}$ for R-MX and $-D^{*}$ for R-XM, and the nonpolar
  reference sits between them.
  (c) The differential transconductance inherits that parity, drawn at the
  subthreshold bias $V_c=0.05$~V. The two
  registries are field-axis mirror images of one another, so they are equal
  and opposite at $D=0$ and not at every field; each crosses zero once, at its
  own compensation field, and the nonpolar reference crosses at $D=0$. Resolving
  the distinct $K$- and $\Gamma$-valley screening moves the calculated
  WSe$_2$ zero from $\Dsym$ by \gdnulllow\% at this bias, too little to see here.}
  \label{fig:landscape}
\end{figure*}

Figure~\ref{fig:landscape} shows the resulting landscape. The two polar registries obey the exact layer-exchange relation
\begin{equation}
\Delta_{K\Gamma}^{\rm MX}(D)=\Delta_{K\Gamma}^{\rm XM}(-D),
\label{eq:mirror}
\end{equation}
for a top--bottom symmetric environment. Polarization moves one R-stacked landscape to opposite positions along the field axis. The nonpolar 2H reference is not exactly a translated member of that family because stacking also renormalizes the layer-symmetric band energies. This distinction matters for quantitative comparison with 2H but not for the registry relation in Eq.~(\ref{eq:mirror}).

The same mirror relation gives an electrical observable. In a symmetric dual gate of the kind routinely built on these materials~\cite{Movva2015_WSe2dualgate}, define the common and differential biases $V_c=(V_t+V_b)/2$ and $V_d=(V_t-V_b)/2$. The differential bias changes the displacement field, and the channel follows: moving the landscape moves the density of states, and with it the hole density.  On a symmetric stack every self-consistent quantity is even about the compensation field.  The drift is even too, and the parity survives.  Equation~(\ref{eq:mirror}) is a statement about the bands; the step to the current needs the device as well: for a top--bottom symmetric stack whose transport does not distinguish the two layers, the same layer-exchange symmetry gives
\begin{equation}
I_{\rm MX}(D)=I_{\rm XM}(-D)
\quad\Longrightarrow\quad
g_d^{\rm MX}(0)=-g_d^{\rm XM}(0),
\label{eq:antisym}
\end{equation}
where $g_d=\partial I/\partial V_d$. The antisymmetry of Eq.~(\ref{eq:antisym}) is exact under those conditions; it relates the two polar states to each other and leaves the magnitude of either open. The two responses can vanish together at a model-dependent common-mode bias, and the sign readout is operated away from that cancellation point. For the centrosymmetric 2H reference, the mirror maps the system to itself and gives $g_d(0)=0$, a control that the same sample can supply: chemical-vapour-deposited bilayers grow 2H and 3R regions on one flake~\cite{DiBerardino2025_2H3R}. The measurement is a small-signal one and asks for low-resistance contacts, which are now available for WSe$_2$~\cite{Shen2021_contact,Pack2024_WSe2}.

A second and complementary readout follows from a second parity, that of each registry about its own compensation field. In the ideal symmetric two-valley model, when the applied field cancels the polar layer asymmetry, the band energies become even about the compensation field. The current is then stationary and $g_d$ crosses zero. That zero is exact only in the ideal symmetric limit: it additionally assumes that transport does not distinguish the layers and that the two valleys share the same compensation field. The direct WSe$_2$ calculations quantify the leading band-structure correction. Fitting each doublet separately on the R-stacked field scan gives $d_\Gamma/d_K=\dratioGK$ and $U^\Gamma/U^K=\UratioGK$, displacing the calculated electrical zero by \gdnullpct\% of $\Dsym$ in strong accumulation and \gdnulllow\% near threshold, where Fig.~\ref{fig:landscape}(c) is drawn. Device asymmetry produces a larger shift: a fractional difference $r$ between the two interfaces displaces that zero by $\mulayerslope\,r$ of $\Dsym$.  A mismatch of one percent already outweighs the band-structure correction.  The device, and not the band structure, limits the practical accuracy of this locator. The built-in field of $\Dbuiltin$~V\,nm$^{-1}$ reported for this system~\cite{Li2026_RWSe2} is the \emph{internal} field $\varphi_0/d_0$, whereas $\Dsym$ is the external one, so the two are not directly comparable. Converted with the screening our own scan implies, which agrees with the out-of-plane response reported for the bilayer~\cite{Laturia2018_dielectric}, they give consistent dipole lengths. The remaining difference lies in the interfacial potential assigned to the $K$ states, which the companion Article analyses~\cite{CompanionArticle}.

This Letter shows what the known $C_3$ selection rule does in a material whose competing valence valleys have opposite interlayer character. In rhombohedral bilayer WSe$_2$, the rule converts ferroelectric layer polarization almost maximally into a $K$--$\Gamma$ energy displacement. Layer exchange then promotes the transduced valley shift to a testable response parity. In a top--bottom symmetric environment the two polarization states give exactly opposite differential responses at zero applied field. In the ideal transport limit a second parity, that of each registry about its own compensation field, pins a zero of that response there, and the valley-dependent screening computed here moves the WSe$_2$ locator by only \gdnullpct\%. More generally, any switchable polar bilayer that combines a symmetry-decoupled valley with a strongly hybridized competing one should transduce in the same way.  These results establish a direct route from stacking symmetry to nonvolatile control and electrical readout of a multivalley landscape.

\begin{acknowledgments}
This work was supported by JSPS KAKENHI (Grants No. JP26H02222, No. JP25K01609, No. JP22H05473, and No. JP21H01019). Numerical calculations were carried out on the Numerical Materials Simulator at the National Institute for Materials Science.
\end{acknowledgments}

\bibliography{references,companion}

\end{document}